\documentclass[prd,twocolumn,showpacs,floats,
nofootinbib,eqsecnum,floatfix]{revtex4-1}

\usepackage{amssymb,amsmath}
\usepackage{epsfig}
\usepackage{color}

\newcommand{\be}{\begin{equation}}
\newcommand{\ee}{\end{equation}}
\newcommand{\ba}{\begin{eqnarray}}
\newcommand{\ea}{\end{eqnarray}}
\newcommand{\non}{\nonumber}

\newcommand{\const}{\mbox{const}}

\newcommand{\eq}[1]{(\ref{#1})}
\newcommand{\n}[1]{\label{#1}}

\newcommand{\hh}{\, ,\hspace{0.8cm}}
\newcommand{\hhh}{\, ,\hspace{0.4cm}}

\begin{document}

\title{Self-energy anomaly of an electric pointlike dipole in three-dimensional static spacetimes}
\author{Valeri P. Frolov}%
\email[]{vfrolov@ualberta.ca}
\affiliation{Theoretical Physics Institute, Department of Physics,
University of Alberta,\\
Edmonton, Alberta, T6G 2E1, Canada
}
\author{Andrey A. Shoom}%
\email[]{ashoom@ualberta.ca}
\affiliation{Theoretical Physics Institute, Department of Physics,
University of Alberta,\\
Edmonton, Alberta, T6G 2E1, Canada
}
\author{Andrei Zelnikov}%
\email[]{zelnikov@ualberta.ca}
\affiliation{Theoretical Physics Institute, Department of Physics,
University of Alberta,\\
Edmonton, Alberta, T6G 2E1, Canada
}


\begin{abstract}
We calculate the self-energy anomaly of a pointlike electric dipole located in a static
$(2+1)$-dimensional curved spacetime. The energy functional
for this problem is invariant under an infinite-dimensional (gauge)
group of transformations parameterized by one scalar function of
two variables. We demonstrate that the problem of the calculation
of the self-energy anomaly for a pointlike dipole can be reduced to the
calculation of quantum fluctuations of an
effective two-dimensional Euclidean quantum field theory. We reduced the problem in question to the calculation of the conformal anomaly of an effective scalar field in two dimensions and obtained an explicit expression for the self-energy
anomaly of an electric dipole in an
asymptotically flat, regular $(2+1)$-dimensional spacetime which may have electrically neutral black-hole-like metrics with regular Killing horizon.

\end{abstract}

\pacs{04.50.Gh, 11.10.Kk, 04.40.Nr \hfill
Alberta-Thy-02-13}

\maketitle

\section{Introduction}

Recently, it was demonstrated that the problem of the self-energy of
pointlike scalar and electric charges in a D-dimensional static gravitational field
can be  reduced to the problem of calculation of vacuum fluctuations
of a scalar field in the
effective (D-1)-Euclidean quantum field theory. This theory, besides
a dynamical scalar field ${\varphi}$ includes (D-1)-dimensional metric
$g_{ab}$ and a dilaton field $\alpha$, which is related to the $g_{tt}$
component of the metric of the original theory. In our previous papers
\cite{FrolovZelnikov:2012a,FrolovZelnikov:2012b} we showed that energy
possesses the property of gauge invariance with respect to joint
transformations of ${\varphi}$, $g_{ab}$, and $\alpha$. Standard
regularizations, required to make the self-energy finite, break this
invariance. As a result, the renormalized expression for the
self-energy for pointlike charges acquires an
anomaly. This anomalous contribution vanishes for even D and it is
nontrivial in odd-dimensional spacetimes. This anomaly was calculated
and discussed in \cite{FrolovZelnikov:2012a,FrolovZelnikov:2012b}. Using this approach we generalized the earlier obtained results \cite{FrolovZelnikov:2012,FrolovZelnikov:2012c,FrolovZelnikov:1982,Lohiya:1982} for the self-energy of charges in the background of static black holes.

In this paper we apply the proposed method to the calculation of self-energy of pointlike electric dipoles.
In three-dimensional spacetime the self-energy of an electric monopole has logarithmic divergency at the spatial infinity. In the case of compact manifolds it results in strong dependence on the boundary conditions.
The anomalous term is local and can be calculated, but it is not very interesting because it boils down to the effective 2D ``cosmological constant" correction. It does not lead to non-trivial self-forces dependent on the space curvature.

On the other hand, the self-energy of a pointlike dipole in three dimensions is infrared (IR) finite and is of the order of ${\bf p}^2/\varepsilon^2$, where ${\bf p}$ is the electric dipole moment of a pointlike particle and $\varepsilon$ is the characteristic size of the particle.
The self-energy anomaly of a dipole also appears in odd-dimensional spacetimes. In the case of a dipole in three dimensions  the anomaly proves to depend on the scalar curvature of a $t=\const$ spatial section. The specific properties of \mbox{$(2+1)$-dimensional} spacetime allow us to calculate the self-energy anomaly  of dipoles explicitly for an arbitrary static spacetime. Here we shall consider regular, asymptotically flat spacetimes which may admit electrically neutral black-hole-like metrics with regular Killing horizon.

An effective three-dimensional electrodynamics often appears in condensed matter systems. The effective 3D Maxwell equations describe, for example, the dynamics of vortices on the films of superfluid ${}^4\mbox{He}$ (see, e.g. \cite{Halperin:1980,Yaremko:2011}).
Usually the effective spatial geometry, which appears in these systems, does not obey the Einstein equations. Therefore, it's important to calculate the self-energy anomaly  in  arbitrary \mbox{(2+1)-dimensional} static geometries.

In the present paper we use the system of units where $G=c=1$ and the sign conventions for the metric
and other geometrical quantities adopted in the books  \cite{MTW,FZ}.

\section{Self-energy of an electric pointlike dipole in a static spacetime}\label{self-energy}

A metric of a static 3-dimensional spacetime can be written in the following form:
\be\label{ds}
ds^2=-\alpha^2 dt^2+g_{ab}\,dx^a dx^b \,,
\ee
where $\alpha$ and $g_{ab}$ are functions of the spatial coordinates $x^{a}, a=1,2$.
The 2-dimensional spatial metric $g_{ab}$, at least locally, can always be written in the conformally flat form
\be\n{cf}
g_{ab}=\Omega^{2}\bar{g}_{ab}\,.
\ee
where $\bar{g}_{ab}$ is the metric of a flat two-dimensional space.  In Cartesian coordinates it reads
\be
\bar{g}_{ab} =\delta_{ab}\,.
\ee
To construct 3-current of a pointlike electric dipole, let us consider two pointlike electric charges $-q$ and $q$
located at the points $x_{0}^{a}$ and $x_{1}^{a}$, respectively. The 3-current of this charge configuration is
\be
J^{\alpha}(x)=\frac{q[\delta^{2}(x-x_{1})-\delta^{2}(x-x_{0})]}{\alpha(x)\sqrt{g(x)}}\delta^{\alpha}_{0}\,,
\ee
where $\delta^{2}(x-x_{0})$ is the 2-dimensional Dirac delta function and $g=\text{det}(g_{ab})$.
Here and in what follows, the Greek indices stand for the spacetime coordinates $(t,x^{a})$.
For $x_{1}^{a}=x_{0}^{a}+\Delta x^{a}$, $\Delta x^{a}\ll 1$, we have
\be
J^{\alpha}(x)\approx\frac{q\Delta x^{a}\partial_{a}\delta^{2}(x-x_{0})}{\alpha(x)\sqrt{g(x)}}\delta^{\alpha}_{0}\,,
\ee
where $\partial_{a}=\partial/\partial x^{a}$. The vector $\Delta x^{a}$ is defined at the point $x_{0}^{a}$ and oriented in the direction from $x_{0}^{a}$ to $x_{1}^{a}$. This vector is tangent to the geodesic defined with
respect to the metric $g_{ab}$ and connecting the points $x_{0}^{a}$ and $x_{1}^{a}$.\footnote{The vector $x_{0}^a$ can, in principle, be defined at any point on the geodesic between $x_{0}^a$ and $x_{1}^a$. In the limit $x_{1}^a\to x_{0}^a$ all such definitions become equivalent.} Thus, one can write
$\Delta x^{a}=|\Delta x^{a}|\,n^{a}$, where $|\Delta x^{a}|$ is the distance between $x_{0}^{a}$ and $x_{1}^{a}$ measured in the metric $g_{ab}$ and $n^{a}$ is a unit vector, $n^{a}n^{b}g_{ab}=1$.

Let us consider the limit
\be
\lim_{\substack{|\Delta x^{a}|\to 0 \\ q\to\infty}}q|\Delta x^{a}|=P\,,
\ee
such that $P$ is a finite quantity. Then, we can define a pointlike dipole moment located at $x_{0}^{a}$ as follows:
\be
p^{a}(x_{0})\equiv Pn^{a}\,.
\ee
The 3-current of the pointlike dipole reads
\be\n{edip}
J^{\alpha}(x)=\frac{p^{a}(x_{0})\partial_{a}\delta^{2}(x-x_{0})}{\alpha(x)\sqrt{g(x)}}\delta^{\alpha}_{0}\,.
\ee
Note that according to our definitions, $P=\Omega\bar{P}$ and $n^{a}=\Omega^{-1}\bar{n}^{a}$, $\Omega>0$, where $\bar{P}$ is defined in the metric $\delta_{ab}$ and $\bar{n}^{a}\bar{n}^{b}\delta_{ab}=1$.
Thus, the pointlike dipole moment
$p^{a}=Pn^{a}=\bar{P}\bar{n}^{a}
=\bar{p}^{a}$ is conformal invariant.

To find the self-energy of any electric charge distribution  we begin with the electromagnetic field action
\be\label{Maxwell}
I=-{1\over 16\pi}\int d^3x\,\alpha\sqrt{g}\,F^{\alpha\beta}F_{\alpha\beta}
+\int d^3x\,\alpha\sqrt{g}\, A_{\alpha }\,J^{\alpha}\,.
\ee
Here $F_{\alpha\beta}\equiv\partial_{\alpha} A_{\beta}-\partial_{\beta} A_{\alpha}$ is the electromagnetic field tensor and $A_{\alpha}$ is the 3-vector potential of the field. The  Maxwell equations are
\be\label{divF}
\nabla_{\beta}F^{\alpha\beta}=4\pi J^{\alpha}\, ,
\ee
where the $\nabla_{\alpha}$ stands for the covariant derivative
defined with respect to the 3-dimensional metric \eq{ds}.
For a static source $J^{\alpha}=\delta^{\alpha}_0 J^0$ we can put $A_{a}=0$, then the
Maxwell equations reduce to the form
\be
\nabla_{\alpha}F^{0\alpha}=4\pi J^{0}\,,
\ee
which can be rewritten in terms of the vector potential $A_{\alpha}=(A_0,0)$ as follows:
\be\label{eqA0}
{1\over \alpha\sqrt{g}}\partial_{a}\left(\sqrt{g}\alpha^{-1}\,g^{ab}
\partial_{b}A_0\right)=4\pi J^0\,.
\ee

Using the static Green function of the Maxwell field \eq{eqA0} one can write
\be\n{A0}
A_0(x)=-4\pi \int_{\Sigma} d^2x'\,\alpha(x')\sqrt{g(x')}\,{\cal G}_{00}(x,x')J^{0}(x')\,,
\ee
where the integral is taken over a 2-dimensional spacelike hypersurface $\Sigma$ and
the Green function ${\cal G}_{00}$ solves the equation
\be\label{eqG00}
{1\over \alpha\sqrt{g}}\partial_{a}\left(\sqrt{g}\alpha^{-1}\,g^{ab}
\partial_{b}{\cal G}_{00}\right)=-{\delta^{2}(x-x')\over \alpha\sqrt{g}}\,.
\ee
The energy of the electromagnetic field associated with the timelike Killing vector $\xi^{\alpha}=\delta^{\alpha}_{0}$
is defined on a spacelike 2-dimensional hypersurface $\Sigma$ as follows:
\be
E\equiv\int_{\Sigma}d^{2}x\sqrt{g}\,T_{\alpha\beta}n^{\alpha}\xi^{\beta}\,,
\ee
where  $n^{\alpha}=\alpha^{-1}\delta^{\alpha}_{0}$ is the unit vector orthogonal to the hypersurface.
The electromagnetic energy-momentum tensor is
\begin{equation}\n{Tem}
T_{\alpha\beta}=\frac{1}{4\pi }\left( F_{\alpha\lambda}\,F_{\beta}{}^{\lambda}
-\frac{1}{4}\,g_{\alpha\beta}\,F_{\lambda\gamma}\,F^{\lambda\gamma}\right)\,,
\end{equation}
and the self-energy of the electric dipole \eq{edip} takes the form
\be\n{E}
E={1\over 8\pi}\int_{\Sigma} d^2x\,\alpha^{-1}\sqrt{g}\,g^{ab}(\partial_{a}A_{0})\,(\partial_{b}A_{0})\,,
\ee
where $A_{0}$ is given by Eq.\eq{A0}.

In what follows, we shall consider asymptotically flat static spacetimes of the form \eq{ds}. Such spacetimes may admit black-hole-like metrics with regular Killing horizons, where $\alpha=0$. Integrating by parts we present the integral in \eq{E} as an integral over the 2-dimensional space and a set of 1-dimensional integrals over the black hole horizon boundaries $l_{h}$ and the boundary at the spatial infinity $l_{\infty}$,
\be\begin{split}\n{E0}
E&=-{1\over 2}\int_{\Sigma} d^2x\,\alpha\sqrt{g}\,A_{0}J^{0}\\
&+{1\over 8\pi}\int_{l_\infty}dl_a \alpha^{-1}\sqrt{g}\,g^{ab} A_0\partial_b A_0\\
&-{1\over 8\pi}\int_{l_h}dl_a \alpha^{-1}\sqrt{g}\,g^{ab} A_0\partial_b A_0\, .
\end{split}\ee
In $2+1$ dimensions, contrary to the case of four and higher-dimensional spacetimes, the self-energy of  pointlike charges diverges at spatial infinity. Nevertheless,  the self-energy of dipoles in low-dimensional electrodynamics is well-defined and IR finite. So, in the case of dipoles the only important divergencies are of ultraviolet (UV) origin.  Here we are focusing on the proper treatment of the UV divergencies for pointlike sources.

For the dipole source in an asymptotically flat $(2+1)$-dimensional spacetime  the vector potential at the spatial infinity decreases as  $A_0\sim |x-x'|^{-1}$, where $|x-x'|$ is the proper distance between the points at $x$ and $x'$. For this reason the surface integrals at $l_{\infty}$ vanish. The potential $A_0$ is constant on a black hole horizon. In this case the surface integrals at $l_{h}$ are proportional to this constant and to the total charge of the black hole, $Q_{h}$,
\be
\int_{l_h}dl_a\alpha^{-1}\sqrt{g}\,g^{ab}\partial_{b}A_0=4\pi \int_{\Sigma} d^2x\,\alpha\sqrt{g}\,J^{0}(x)=4\pi Q_h\,.
\ee
We consider the problem when the only charges are related to dipole in question and all black holes are neutral, i.e., $Q_h=0$. In this case all the surface integrals at $l_{h}$ vanish as well. Then, using the expression \eq{A0} we present Eq.\eq{E0} in the form \cite{FrolovZelnikov:2012b}
\ba\label{E1}
E&=&2\pi\int_{\Sigma} d^{2}x\,d^{2}x'\,\alpha(x)\sqrt{g(x)}\alpha(x')\sqrt{g(x')}\non\\
&\times&J^{0}(x)\,{\cal G}_{00}(x,x')\,J^{0}(x')\,.
\ea

Following to the lines of the paper \cite{FrolovZelnikov:2012b}
we introduce a new field variable $\psi$ such that
\be\n{Apsi}
A_0=-\alpha^{1/2}\,\psi\,.
\ee
which satisfies the equation
\be\begin{split}\label{eqpsi}
{\cal O}\,\psi&=-4\pi j\hh
{\cal O} \equiv \triangle  +V\,.
\end{split}\ee
Here $\triangle$ is the 2-dimensional Laplace-Beltrami operator
defined with respect to the metric $g_{ab}$,
$V$ is the potential, and $j$ is the effective scalar charge density,
\be\label{Vj}
V=-\alpha^{1/2}\triangle(\alpha^{-1/2})\hh j\equiv\alpha^{3/2} J^0\, .
\ee
The field $\psi$ is chosen in such a way that the operator
$\cal O$ is self-adjoint in the space with the metric $g_{ab}$.

In terms of the Green function of the operator \eq{eqpsi}  we obtain
\be\n{psi}
\psi(x)=4\pi\int_{\Sigma} d^{2}x'\,\sqrt{g(x')}\,{\cal G}(x,x')\,j(x')\,.
\ee
The Green function ${\cal G}$ solves the following equation:
\be\n{eqG}
(\triangle+V){\cal G}(x,x')=-\frac{\delta^{2}(x-x')}{\sqrt{g}}\,.
\ee
The Green functions ${\cal G}$ and ${\cal G}_{00}$ are related to each other as follows:
\be
{\cal G}_{00}(x,x')=\alpha^{1/2}(x)\,\alpha^{1/2}(x')\,{\cal G}(x,x')\,
\ee
and the energy functional \eq{E1} can be written as
\be\begin{split}\label{E2}
E&=2\pi\int_{\Sigma} d^{2}x\,d^{2}x'\,\sqrt{g(x)}\sqrt{g(x')}\,j(x)\,{\cal G}(x,x')\,j(x')\, .
\end{split}\ee

Similarly to the case of electric charges in higher-dimension \cite{FrolovZelnikov:2012b} the self-energy functional for the energy of charge distributions in $2+1$ dimensions is invariant under the following transformations of the metric \eq{ds},  the field $\psi$, and the effective scalar charge density $j$:
\be\begin{split}\label{trans}
&g_{ab}=\Omega^2 \bar{g}_{ab}\hhh \alpha=\bar{\alpha}\hhh
\psi=\bar{\psi}\hhh j=\Omega^{-2}\bar{j}\,.
\end{split}\ee

From the point of view of a field theory on a 2-dimensional spatial slice,
this symmetry is just the symmetry under the conformal transformation of the 2-dimensional metric
$g_{ab}$. Applying these transformations to the operator ${\cal O}$ we derive
\be
{\cal O}\psi=(\triangle  +V)\,\psi = \Omega^{-2}(\bar{\triangle}+\bar{V})\,\bar{\psi}\,,
\ee
where $\bar{\triangle}$ is the 2-dimensional Laplace-Beltrami operator
defined with respect to the flat metric  $\bar{g}_{ab}$.
Note that the symmetry transformations \eq{trans} are identical to those for the self-energy of a scalar charge
\cite{FrolovZelnikov:2012a}. This identity happens to be true in a 3-dimensional spacetime only.
The invariance with respect to the transformations \eq{trans} describes the
classical symmetry of the system.

For a pointlike charge distributions, the
classical functional \eq{E} diverges. The divergent part of the electromagnetic
energy  can be recombined with the contribution of
nonelectromagnetic fields, which are responsible for the stability of a
charge distribution (in our case a dipole), and also contribute to its bare mass. After this renormalization one
obtains the finite total mass.
Note that in a generic case nonelectromagnetic fields do not respect the observed symmetry.
This is the cause of an anomalous
contribution to the self-energy of charge distributions in curved spacetimes. To deal with the divergencies and to extract the anomalous terms we apply the regularization methods of quantum field theory  \cite{FrolovZelnikov:2012a,FrolovZelnikov:2012b,FrolovZelnikov:2012c}.
In quantum field theory the fact that
renormalization procedure breaks some symmetries of the classical theory is the cause of appearance of conformal,
chiral, and other anomalies. In our case,  the same arguments are applicable to
the renormalized self-energy of classical sources. For the same reason their
self-energy acquires anomalous terms.  In fact, one could try to regularize self-energy of charges in quantum field theory from the very beginning and then apply the result to static charges,  or, the other way around, apply quantum field theory regularization methods to classical static charge distributions. Both methods can be reduced to the regularization procedures of the Green functions in D and (D-1) dimensions correspondingly. As it was demonstrated in \cite{CasalsPoissonVega:2012} these two regularization schemes eventually lead to the same result. In higher dimensions the structure of divergencies becomes considerably more complicated than in D$\le 4$ (see, e.g., \cite{GaltsovSpirin:2007}).

All the traditional methods of UV regularization
like point-splitting, zeta-function
and dimensional regularizations, proper time cutoff,  Pauli-Villars, and other
approaches are applicable to the calculation of the self-energy. For our problem
the most natural choice is to use the point-splitting regularization.

\section{Renormalized self-energy of an electric pointlike dipole}

Using the expression \eq{edip} for the 3-current $J^\alpha$ and the definition \eq{Vj} for $j(x)$ the self-energy of the pointlike
dipole \eq{E2} can be written as
\ba\n{ef}
E(x_{0})&=&2\pi\int_{\Sigma} d^{2}x\,d^{2}x'\,\delta^{2}(x-x_{0})\delta^{2}(x'-x_{0})\n{E3}\\
&\times&p^{a}(x_{0})p^{b'}(x_{0})[\alpha^{1/2}(x)\alpha^{1/2}(x'){\cal G}(x,x')]_{;ab'}\non\,.
\ea
Here, in accordance with our previous discussion, we preformed the integration by parts and ignored vanishing surface terms. Here and in what follows, the semicolon
stands for the covariant derivative defined with respect to the metric $g_{ab}$.

The self-energy diverges due to pointlike nature of electric charges. In analogy with the calculation of the self-energy of an electric charge, we shall calculate the renormalized self-energy of the electric dipole as follows:
\ba
E_{\text{ren}}(x_{0})&=&2\pi\int_{\Sigma} d^{2}x\,d^{2}x'\,\delta^{2}(x-x_{0})\delta^{2}(x'-x_{0})\n{E4}\\
&\times&p^{a}(x_{0})p^{b'}(x_{0})[\alpha^{1/2}(x)\alpha^{1/2}(x'){\cal G}(x,x')_{\text{reg}}]_{;ab'}\non\,,
\ea
where
\ba
&&[\alpha^{1/2}(x)\alpha^{1/2}(x'){\cal G}_{\text{reg}}]_{;ab'}=\frac{(\partial_{a}\alpha(x))(\partial_{b'}\alpha(x'))}{4\alpha^{1/2}(x)\alpha^{1/2}(x')}{\cal G}_{\text{reg}}\non\\
&&+\frac{\alpha^{1/2}(x)}{2\alpha^{1/2}(x')}(\partial_{b'}\alpha(x'))\partial_{a}{\cal G}_{\text{reg}}
+\frac{\alpha^{1/2}(x')}{2\alpha^{1/2}(x)}(\partial_{a}\alpha(x))\partial_{b'}{\cal G}_{\text{reg}}\non\\
&&+\alpha^{1/2}(x)\alpha^{1/2}(x'){\cal G}_{\text{reg};ab'}\,,\n{E4a}
\ea
where for brevity we have dropped the arguments of ${\cal G}$.
Here ${\cal G}_{\text{reg}}$ is the regularized Green function,
\be\n{Greg}
{\cal G}_{\text{reg}}={\cal G}-{\cal G}_{\text{div}}\,,
\ee
and the diverging part ${\cal G}_{\text{div}}$ contains all the diverging terms appearing in ${\cal G}$, $\partial_{a}{\cal G}$, $\partial_{b'}{\cal G}$, and ${\cal G}_{;ab'}$.

It should be noted that, in terms of the field $\psi$ \eq{Apsi}  the self-energy \eq{E}
\be\label{Eem}
E={1\over 8\pi}\int_{\Sigma} d^2x\,\sqrt{g}\,g^{ab}\left(\partial_{a}\psi+{\partial_{a}\alpha\over 2\alpha}\psi\right)\,\left(\partial_{b}\psi+{\partial_{b}\alpha\over 2\alpha}\psi\right)\,,
\ee
has the form of a two-dimensional Euclidean action functional
\be\label{IE}
I_E={1\over 8\pi}\int_{\Sigma} d^2x\,\sqrt{g}\,\psi{\cal O}\psi+\mbox{boundary terms}\,.
\ee
The functional \eq{Eem} is similar to the functional Eq.(2.7) of the paper \cite{FrolovZelnikov:2012a}, where we have studied the scalar charge distributions. The only difference is how interaction with the ``dilaton" $\alpha$ enters the operator ${\cal O}$.

One can quantize this effective two-dimensional  Euclidean field theory \eq{IE}. The quantum average $ \left<\psi(x)\psi(x')\right>$ over
the Euclidean vacuum state  $|0\rangle$ gives the Euclidean Green function  ${\cal G}(x,x')$
\be
\left<\psi(x)\psi(x')\right>={\cal G}(x,x')\,,
\ee
which corresponds to the operator ${\cal O}$ (see \eq{eqpsi},\eq{eqG})
\be
{\cal O}\,{\cal G}(x,x')=-\delta^2(x,x') \,.
\ee
Using this relation between the Green function and the quantized field $\psi$ and subtracting divergencies we get
\be
{\cal G}_{\text{reg}}(x,x')=\left<\psi(x)\psi(x')\right>_{\text{ren}}\,.
\ee
Then we can present the self-energy as follows:
\ba\n{eren}
E_{\text{ren}}(x_0)&=&2\pi\bigg[\frac{1}{4\alpha}p^{a}p^{b'}\alpha_{;a}\alpha_{;b'}
\left<\psi^2\right>_{\text{ren}}\non\\
&+&p^a\alpha_{;a}\left<\psi\, p^{b'}\psi_{;b'}\right>_{\text{ren}}\non\\
&+&\alpha\left<p^{a}\psi_{;a}\,
p^{b'}\psi_{;b'}\right>_{\text{ren}}
\bigg]\biggr\rvert_{(x,x')\to x_0}\,.
\ea
In this representation, one can use the regularization methods of quantum field theory.

According to the expressions \eq{edip} and \eq{trans}, the Green function and the directional derivative $p^a\partial_a$ are conformal invariant. Therefore, the nonrenorlmalized $\left<\psi^2\right>$, $\left<p^a\psi_{;a}\psi\right>$, and $\left<p^a\psi_{;a}p^b\psi_{;b}\right>$ are conformal invariant as well. If we subtract the diverging parts of these expressions, in accordance with \eq{Greg}, this conformal invariance gets broken. In order to restore the invariance, one should add an anomalous terms $A(x)$, such that
\ba\n{ins}
&&\left<\psi^2\right>_{\text{ren}}+A_1=
\left<\bar{\psi}^2\right>_{\text{ren}}
+\bar{A}_1=\text{inv}\,,\non\\
&&\left<p^a\psi_{;a}
\psi\right>_{\text{ren}}+A_2=
\left<p^a\bar{\psi}_{;a}
\bar{\psi}\right>_{\text{ren}}
+\bar{A}_2=\text{inv}\,,\\
&&\left<p^a\psi_{;a}
p^b\psi_{;b}\right>_{\text{ren}}+A_3=
\left<p^a\bar{\psi}_{;a}
p^b\bar{\psi}_{;b}\right>_{\text{ren}}
+\bar{A}_3=\text{inv}\,.\non
\ea
The differences between the anomalous terms
\ba\n{andif}
&&B_1=A_1-\bar{A}_1=
\left<\bar{\psi}^2\right>_{\text{ren}}-
\left<\psi^2\right>_{\text{ren}}\,,\non\\
&&B_2=A_2-\bar{A}_2=
\left<p^a\bar{\psi}_{;a}
\bar{\psi}\right>_{\text{ren}}-
\left<p^a\psi_{;a}
\psi\right>_{\text{ren}}\,,\\
&&B_3=A_3-\bar{A}_3=
\left<p^a\bar{\psi}_{;a}
p^b\bar{\psi}_{;b}\right>_{\text{ren}}-
\left<p^a\psi_{;a}p^b
\psi_{;b}\right>_{\text{ren}}\,,\non
\ea
can be calculated as follows:
\ba
B_1&=&\bar{\cal G}_{\text{reg}}-
{\cal G}_{\text{reg}}=({\cal G}_{\text{div}}
-\bar{\cal G}_{\text{div}})|_{x'\to x}\,,\non\\
B_2&=&p^a(\bar{\cal G}_{\text{reg};a}-{\cal G}_{\text{reg};a})=p^{a}({\cal G}_{\text{div};a}
-\bar{\cal G}_{\text{div};a})|_{x'\to x}\,,\non\\
B_3&=&p^ap^{b'}(\bar{\cal G}_{\text{reg};ab'}-{\cal G}_{\text{reg};ab'})\\
&=&p^{a}p^{b'}({\cal G}_{\text{div};ab'}-\bar{\cal G}_{\text{div};ab'})|_{x'\to x}\,.\non
\ea

\subsection{The Schwinger--DeWitt expansion of the Green function}

To calculate the diverging part ${\cal G}_{\text{div}}$ we use Schwinger--DeWitt expansion. We begin with the heat kernel expansion for the operator ${\cal O}$ [see Eq.\eq{eqpsi}] and define the heat kernel ${\cal K}(s|x,x')$ as a solution of the equation
\be\begin{split}
\left[-{\partial\over \partial
s}+{\cal O} -m^{2}\right]\,{\cal K}(s|x,x')
&=-\frac{\delta^{2}(x-x')}{\sqrt{g}}\,\delta(s)\, ,
\end{split}\ee
where we introduced the mass term $m^{2}$. Then, the Green function ${\cal G}$ is calculated as [cf. Eq.\eq{eqG}]
\be\label{calG}
{\cal G}(x,x')=\lim_{m^{2}\to0}\int_0^\infty ds\,{\cal K}(s|x,x')\,.
\ee
The divergent terms of ${\cal G}$ which define ${\cal G}_{\text{div}}$ are determined by
the behavior of the heat kernel at sufficiently small $s$
and can be found by using the standard Schwinger--DeWitt expansion
\ba\n{calK}
{\cal K}(s|x,x')&=&{\Delta^{1/2}(x,x')\over 4\pi
s}\,\exp\left({-{{\sigma}(x,x')\over 2 s}}-m^{2}s\right)\,\non\\
&\times &\sum_{k=0}^{\infty}
{a}_k(x,x')s^k\,,
\ea
where $\Delta^{1/2}(x,x')$ is the Van Vleck--Morette determinant,
which satisfies the following relation $(\sigma_{;a}\equiv\sigma_{a})$:
\be
2\Delta^{1/2}=2\Delta^{1/2}\!{}_{;a}\sigma^{a}+\Delta^{1/2}\sigma^{a}\!{}_{;a}\,,
\ee
$\sigma=(1/2)\sigma^{a}\sigma_{a}$ is one-half of the square of the proper distance between $x$ and $x'$ measured in the metric $g_{ab}$, and ${a}_k$'s  are the Schwinger--DeWitt coefficients
for the operator ${\cal O}$. In the limit $m^{2}\to0$ they satisfy the following relation:
\be
\sigma^{a}a_{k+1;a}+(k+1)a_{k+1}=\frac{(\Delta^{1/2}a_{k})^{;a}_{\,\,\,;a}}{\Delta^{1/2}}+Va_{k}\,,
\ee
where $a_{0}=1$.

Taking the limit $m^{2}\to0$ and neglecting the mass scale terms $\ln(m^{2})$ we derive the sought diverging part of the Green function
\be
{\cal G}_{\text{div}}=\frac{\Delta^{1/2}}{8\pi}\left[(a_{1}\sigma-2)(\ln(\sigma/2)+2\gamma)-a_{1}\sigma\right]\,,
\ee
where $\gamma\approx0.57722$ is Euler's constant. To calculate ${\cal G}$, $\partial_{a}{\cal G}$, $\partial_{b'}{\cal G}$, and ${\cal G}_{;ab'}$ in the coincidence limit $x'\to x$ $(\sigma\to0)$ we use the covariant expansions of $\Delta^{1/2}$, $a_{1}$, $\sigma_{;b'}$, and $\sigma_{;ab'}$ (see \cite{Christen1,Christen2}). The expansions of the sufficient order are given in the Appendix. Keeping the diverging and zero-order terms we derive
\ba
{\cal G}_{\text{div}}&=&-{1\over4\pi}\left(\ln(\sigma/2)+2\gamma\right)\,,\n{Gab}\\
{\cal G}_{\text{div};a}&=&-{\sigma_{a}\over4\pi\sigma}\hh {\cal G}_{\text{div};b'}\to
g_{b}^{\,\,b'}{\cal G}_{\text{div};b'}={\sigma_{b}\over4\pi\sigma}\,,\non\\
{\cal G}_{\text{div};ab'}&\to& g_{b}^{\,\,b'}{\cal G}_{\text{div};ab'}={1\over4\pi}\left[{g_{ab}\over\sigma}-{\sigma_{a}\sigma_{b}\over\sigma^{2}}-{R\sigma_{a}\sigma_{b}\over12\sigma}\right.\non\\
&+&\left.{Rg_{ab}\over4}-{V\sigma_{a}\sigma_{b}\over2\sigma}-{Vg_{ab}\over2}
\left(\ln(\sigma/2)+2\gamma\right)\right]\non\,.
\ea

\section{Self-energy anomaly}

The nonrenormalized self-energy functional \eq{ef} is  invariant under spatial conformal transformations \eq{trans}. According to the expressions \eq{eren} and \eq{ins}, a subtraction of its diverging part breaks the invariance. There is the corresponding self-energy anomaly term which allows to restore the broken invariance as follows:
\be
E_{\text{ren}}+A=
\bar{E}_{\text{ren}}+\bar{A}=\text{inv}\,.
\ee
In accordance with \eq{eren} and \eq{andif}, we can define the self-energy anomaly, $\Delta E_{\text{ren}}$, as
\ba
\Delta E_{\text{ren}}&\equiv& A-\bar{A}=
\bar{E}_{\text{ren}}-E_{\text{ren}}\non\\
&=&
2\pi\bigg[\frac{(p^a\alpha_{;a})^2}{4\alpha}B_1+p^a\alpha_{;a}B_2+\alpha
B_3\bigg]\,.
\ea
To calculate the self-energy anomaly we use the relations between $\sigma$ and $\bar{\sigma}$ in the limit $x'\to x$ (see Appendix A of \cite{FrolovZelnikov:2012c})
\be
\sigma\approx\Omega^{2}\bar{\sigma}\hh \sigma_{a}\approx\Omega^{2}
\bar{\sigma}_{a}\,.
\ee
Evaluating the integral\footnote{
The delta functions in this integral should be understood as the limit of a smeared localized distribution. For example one can consider $\delta^{2}(x-x_{0})\rightarrow \delta(|x-x_0|^2-\epsilon^2)/\pi$, which is a distribution
 over a circle of a small radius $\epsilon$. Performing integration at first and then taking the limit $\epsilon\rightarrow 0$ one can reproduce the formula \eq{deltaab}.
}
\be\label{deltaab}
\int_{\Sigma} d^{2}x\,d^{2}x'\delta^{2}(x-x_{0})\delta^{2}(x'-x_{0}){\bar{\sigma}_{a}\bar{\sigma}^{b}\over\bar{\sigma}}=\delta_a^b
\ee
we derive from Eq.\eq{Gab}
\ba
B_1&=&-\frac{1}{2\pi}\ln(\Omega)\hhh
B_2=0\,,\non\\
B_3&=&p^ap^bg_{ab}\frac{1}{4\pi}\left(\frac{R}{6}-V\ln(\Omega)\right)\,.
\ea
Then, the self-energy anomaly is
\ba\n{AE}
\Delta E_{\text{ren}}&=&
\frac{\alpha}{12}R p^ap_a-\left[\frac{(p^a\alpha_{;a})^2}{4\alpha}+\frac{\alpha}{2}p^ap_aV\right]\ln(\Omega)\,,
\non\\
\ea
where the Ricci scalar $R$ corresponding to the metric $g_{ab}$ is expressed as follows:
\be
R={2\over\Omega^{4}}[\Omega^{;a}\Omega_{;a}-\Omega\Omega^{;a}_{\,\,;a}]\,.
\ee

Ultrastatic 3D spacetimes are especially interesting, since they naturally appear as effective geometries for excitations in a variety of condensed matter systems. Using the expression \eq{AE} one can easily calculate the self-energy of a pointlike dipole in an ultrastatic spacetime $\alpha=1$,
\be
E_{\text{ren}}=-{R\over12}p^{a}p_{a}\,.
\ee

\section{Conclusions}

In this paper, we demonstrated on the example of a three-dimensional static spacetime that the renormalized self-energy of a pointlike electric dipole in a static odd-dimensional spacetime
contains an anomaly contribution. Note that this is quite general property and within our considerations we did not require that a three-dimensional static spacetime satisfies the Einstein equations. We calculated explicitly this anomaly in a three-dimensional static spacetime. In ultrastatic spacetimes the anomaly accounts for the whole effect for the self-energy (self-mass) correction of an electric dipole. In this case the self-energy correction proves to depend on the geometry of the two-dimensional space and is proportional to its scalar curvature.

It is remarkable that a solution of the rather old classical problem of the self-energy can be reduced to the problem of Euclidean quantum field theory in the space of the codimension one. Using the methods of quantum field theory we are able to perform calculations of the classical self-energy of a very generic setup of the system of charges.

On can expect that  the renormalized value of the
self-energy does not depend on the details of the regularization scheme \cite{Moretti:1999}. In $2+1$ dimensions the problem of self-energy is reduced to the effective quantum theory in a genetic two-dimensional space. But any two-dimensional space is conformally flat and the transformations \eq{trans} relate an ultrastatic $(2+1)$-dimensional spacetime to flat one. This is why one can obtain an answer for the dipole self-energy explicitly for generic ultrastatic $(2+1)$-dimensional spacetimes.

\acknowledgments

This work was partly supported  by  the Natural Sciences and Engineering
Research Council of Canada. The authors are also grateful to the
Killam Trust for its financial support.

\appendix

\section{Covariant  expansions of the bitensor quantities}

Covariant expansions of bitensor quantities defined at the points $x$ and $x'$
in terms of functions at $x$ and the tangent vector $\sigma^{a}\equiv\sigma^{;a}$
are done by transforming them into tensor quantities at $x$, with the aid of the bivector $g_{a}^{\,\,a'}$
and the subsequent expansion (for details see, e.g., \cite{Christen1,Christen2})
\ba
{\sigma}_{;ab'}&\to& g_{b}^{\,\,b'}{\sigma}_{;ab'}=-g_{ab}-\frac{1}{6}R_{acbd}\sigma^{c}\sigma^{d}\non\\
&+&{1 \over 12}R_{acbd;e}\sigma^{c}\sigma^{d}\sigma^{e}+...\,,\non\\
\Delta^{1/2}&=&1+{1 \over 12}R_{ab}\sigma^{a}\sigma^{b}
-{1 \over 24}R_{ab;c}\sigma^{a}\sigma^{b}\sigma^{c}+...\,,\non\\
{\Delta^{1/2}}_{;a}&=&{1\over6}R_{ab}\sigma^{b}-{1\over24}(2R_{ab;c}-R_{bc;a})\sigma^{c}\sigma^{b}+...\,,\non\\
{\Delta^{1/2}}_{;b'}&\to&g_{b}^{\,\,b'}{\Delta^{1/2}}_{;b'}=-{1\over6}R_{bc}\sigma^{c}\non\\
&+&{1\over24}(2R_{bc;d}+R_{cd;b})\sigma^{c}\sigma^{d}+...\,,\n{A1}\\
{\Delta^{1/2}}_{;ab'}&\to&g_{b}^{\,\,b'}{\Delta^{1/2}}_{;ab'}=-{1\over6}R_{ab}\non\\
&+&{1\over12}(R_{ab;c}+R_{ac;b}-R_{bc;a})\sigma^{c}+...\,,\non\\
a_{1}&=&{1\over6}R+V-{1\over12}R_{;a}\sigma^{a}-{1\over2}V_{;a}\sigma^{a}+...\,,\non\\
a_{1;a}&=&{1\over12}R_{;a}+{1\over2}V_{;a}+...\,,\non\\
a_{1;b'}&\to&g_{b}^{\,\,b'}a_{1;b'}={1\over12}R_{;b}+{1\over2}V_{;b}+...\non\,.
\ea
Here $R_{abcd}$, $R_{ab}$, and $R$ are the Riemann tensor, the Ricci tensor, and the Ricci scalar defined
with respect to the 2-dimensional metric $g_{ab}$. Note that in a 2-dimensional space (see, e.g., \cite{FZ})
\ba
R_{abcd}&=&{1\over2}(g_{ac}g_{bd}-g_{ad}g_{bc})R\,,\non\\
R_{ab}&=&{1\over2}g_{ab}R\n{A2}\,.
\ea

\end{document}